# Solving the Schrödinger equation with genetic algorithms: a practical approach.

Rafael Lahoz-Beltra [1,2]

[1] Department of Biodiversity, Ecology and Evolution (Biomathematics), Faculty of Biological Sciences, Complutense University of Madrid, 28040 Madrid, Spain; lahozraf@ucm.es

[2] Modeling, Data Analysis and Computational Tools for Biology Research Group, Complutense University of Madrid, 28040 Madrid, Spain

**Abstract:** The Schrödinger equation is one of the most important equations in physics and chemistry and can be solved in the simplest cases by computer numerical methods. Since the beginning of the 70's of the last century the computer began to be used to solve this equation in elementary quantum systems, e.g. and in the most complex case a 'hydrogen-like' system. Obtaining the solution means finding the wave function, which allows predicting the physical and chemical properties of the quantum system. However, when a quantum system is more complex than a 'hydrogen-like' system then we must be satisfied with an approximate solution of the equation. During the last decade the application of algorithms and principles of quantum computation in disciplines other than physics and chemistry, such as biology and artificial intelligence, has led to the search for alternative techniques with which to obtain approximate solutions of the Schrödinger equation. In this paper, we review and illustrate the application of genetic algorithms, i.e. stochastic optimization procedures inspired by Darwinian evolution, in elementary quantum systems and in quantum models of artificial intelligence. In this last field, we illustrate with two 'toy models' how to solve the Schrödinger equation in an elementary model of a quantum neuron and in the synthesis of quantum circuits controlling the behavior of a Braitenberg vehicle.

**Keywords:** Schrodinger's equation; Wave function; Genetic algorithms; Quantum computing; Quantum artificial intelligence; Quantum neuron; Quantum Braitenberg vehicle.

## 1. Introduction

One of the central problems of quantum mechanics entails in solving the Schrödinger equation. This equation, one of the most important in physics, describes the energy and temporal evolution of a quantum system as a set of the different quantum states or energy levels in which a mechanical system can be found. Thus, the evolution of an isolated quantum system from an initial state to another final state, e.g. a system comprising a particle, is ruled by the equation:

$$i\hbar \frac{\partial}{\partial t}|\psi(t)\rangle = H(t)|\psi(t)\rangle \quad (1)$$

where $i$ is the imaginary number, $\hbar$ is a constant named "reduced Planck constant" and $H(t)$ is known as the "Hamiltonian operator". In the equation $|\psi(t)\rangle$ is the time-dependent state vector describing the state of the quantum system at time $t$.

In the simplest cases, the Schrödinger equation is solved by numerical methods treating it as a second order differential equation. A common practice is to separate the time-dependent portion of the wavefunction from the spatial term of the equation, yielding as a result of this step a time-independent form of the Schrödinger equation.

Therefore, for the simplest quantum systems, that is, those that are one-dimensional and independent of time, we will write the Schrödinger equation as follows:

$$\widehat{H}|\psi\rangle = E|\psi\rangle \quad (2)$$

where $\widehat{H}$ is the Hamiltonian operator of the system under study, $E$ the total energy and $|\psi\rangle$ the wave function. Solving the Schrödinger equation involves finding the wave function $|\psi\rangle$, which in practice is of great convenience. For instance, in quantum chemistry if we know the wave function then we can predict the physical and chemical properties of the molecules. Thus, we can find out the shape and orientation of the orbitals of the electrons in an atom or molecule. We can also study chemical bonds or build a model of a crystal. However, in practice, solving the equation analytically is only possible for very simple cases, e.g., when the quantum system is 'hydrogen-like'. In other words, the quantum system is composed by a nucleus with one electron such as H, $He^+$, etc. In all other circumstances, i.e. when the quantum system is more complex, we will be satisfied with solving the Schrödinger equation in an approximate way. Whether for scientific or educational purposes, the search for a numerical solution of the Schrödinger equation of a physical system has always been a subject of study. In 1972 [1] proposed to obtain the solution of the equation with a computer by applying the Numerov method, illustrating the application of the algorithm with a simple quantum harmonic oscillator.

Computer numerical techniques are nowadays commonly used, obtaining an approximate solution of the Schrödinger equation by means of different methods. For instance, using a computer algebra system like Maxima [2] it is possible to solve the Schrödinger equation for a simple quantum system, as is the case of a particle in a box. Classical numerical methods such as the Runge-Kutta method have also been used to solve the Schrödinger equation in 'two body systems' e.g. the hydrogen atom [3]. The efficiency of the method has been compared with other methods of numerical integration. For example, [4] solved the Schrödinger equation for the propagation of a Gaussian wave packet based on Euler, Runge-Kutta of second and fourth orders algorithms, etc.

Other common approach consists of applying the finite difference method [5] finding the solution of the equation by transforming the differential equation into a matrix equation:

$$\begin{pmatrix} U(x_1) & 0 & \ldots & \ldots & \ldots & 0 \\ 0 & U(x_2) & 0 & \ldots & \ldots & \vdots \\ \vdots & 0 & \ddots & & & \vdots \\ \vdots & \vdots & & \ddots & & \vdots \\ \vdots & \vdots & & & \ddots & 0 \\ 0 & \ldots & \ldots & \ldots & 0 & U(x_N) \end{pmatrix} \begin{pmatrix} \psi_1 \\ \psi_2 \\ \vdots \\ \vdots \\ \vdots \\ \psi_N \end{pmatrix} = E \begin{pmatrix} \psi_1 \\ \psi_2 \\ \vdots \\ \vdots \\ \vdots \\ \psi_N \end{pmatrix} \quad (3)$$

where the above matrix is the Hamiltonian matrix. Note that the Hamiltonian matrix involves the sum of two matrices. Thus, in (4) the left matrix represents the kinetic energy and the right matrix the potential energy. In the right matrix $V(x)$ is the potential energy, the Schrödinger equation being linear when the potential energy is independent of the wave function. Likewise, the left matrix represents the kinetic energy, which is approximated by applying the finite difference method. The diagonalization of the Hamiltonian, from resulting sum of these two matrices will lead to obtain the energies and wave functions. As an example, in the simple harmonic oscillator if we study four points of a region, we will get that the following matrix equation:

$$-\frac{1}{2h^2}\begin{pmatrix} -2 & 1 & 0 & 0 \\ 1 & -2 & 1 & 0 \\ 0 & 1 & -2 & 1 \\ 0 & 0 & 1 & -2 \end{pmatrix}\begin{pmatrix} \psi_1 \\ \psi_2 \\ \psi_3 \\ \psi_4 \end{pmatrix} + \begin{pmatrix} V_1 & 0 & 0 & 0 \\ 0 & V_2 & 0 & 0 \\ 0 & 0 & V_3 & 0 \\ 0 & 0 & 0 & V_4 \end{pmatrix}\begin{pmatrix} \psi_1 \\ \psi_2 \\ \psi_3 \\ \psi_4 \end{pmatrix} = E\begin{pmatrix} \psi_1 \\ \psi_2 \\ \psi_3 \\ \psi_4 \end{pmatrix} \quad (4)$$

where the parameter $h$ of the Hamiltonian is the step size. This approach is shown by [6] in an elementary quantum system such as the quantum harmonic oscillator, illustrating with a simple code in the Python language how it is possible to diagonalize the Hamiltonian matrix. The result of these operations will be the eigenvalues (energies $E$) and eigenvectors (wave functions $|\psi\rangle$).

Using also the Python language there is described a more sophisticated option oriented to education [7]. It consists of an open source virtual laboratory with which it is possible to approach the numerical simulation of both the linear and the nonlinear Schrödinger equation dependent on time. When the potential energy depends on the wave function then the Schrödinger equation is nonlinear. This situation of great interest in disciplines as varied as biology, cosmology or optics. In this case, the authors of [7] applied the method known as split-step Fourier or beam propagation method. Using the Crank-Nicolson numerical method, [8] solves the time-dependent Schrödinger equation with visual programming in LabVIEW. In the scientific literature, there are also some interesting approaches such as the case of [9]. Using an analog computer the Schrödinger equation is solved in a one-dimensional time-independent quantum system.

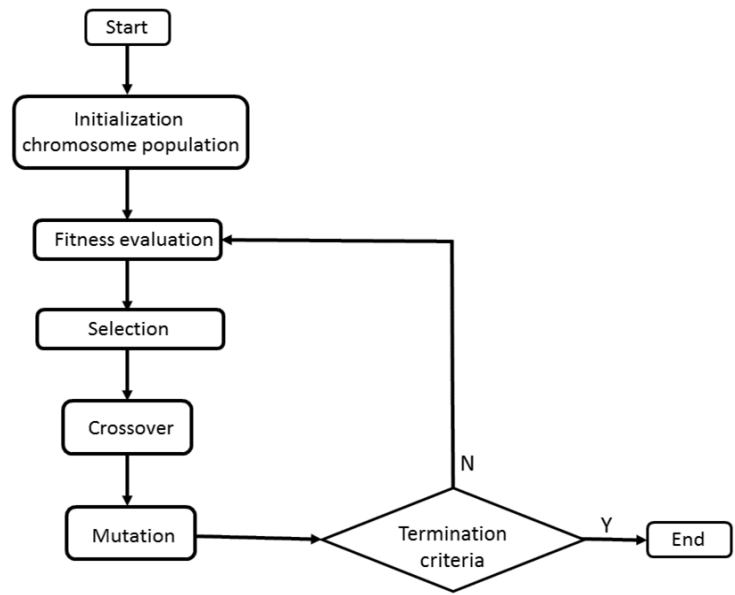

**Figure 1**. Flowchart of a standard genetic algorithm (SGA).

During the last decade the application of some algorithms and principles of quantum computation in disciplines other than physics and chemistry, such is the case of biology and artificial intelligence, has made it necessary to explore other methods to obtain approximate solutions of the Schrödinger equation. One of the most interesting approaches has been the application of evolutionary computation methods, such as genetic algorithms. Genetic algorithms (GA) are stochastic optimization procedures inspired by the theory of natural selection [10, 11, 12] introduced by Charles Darwin in 1859. They are currently used in optimization problems in various disciplines [11], since they are methods for exploring and evaluating a space of solutions until an optimal solution, either global or local, is found.

## 2. Genetic algorithms in quantum protocols

A GA generally comprises the following steps (Figure 1). First, (a) a population of solutions is randomly generated. The solutions are vectors representing chromosomes, with substrings representing genes. These vectors may contain 0's and 1's, integers, real numbers and so on. Next, (b) goodness or validity of the solutions is evaluated. This goodness referred to as fitness represents in Nature a measure of an individual's adaptation to the environment. Indeed, in an optimization problem, the fitness evaluates the goodness of a solution in a certain environment, which is the problem space. Solutions with a higher fitness value will have a higher probability of surviving, and consequently (c) of reproducing, being selected and passing on to the next generation. Therefore, those solutions or chromosome with a higher fitness will have a larger probability of propagation of the genes to the next generation. Now, the individuals, chromosomes or solutions passed on to the next generation will be modified by genetic mechanisms. These mechanisms are mainly two: (d) mutation and/or cross over (Figure 2), giving rise to a new generation of individuals, namely the offspring, which will be genetically different from the parental generation from which they originate. The search process shown in Figure 2 is repeated over a certain number of generations, ending when the termination criterion is achieved. For instance, the optimal or near-optimal solution or chromosome is found, or when a certain maximum number of generations is reached.

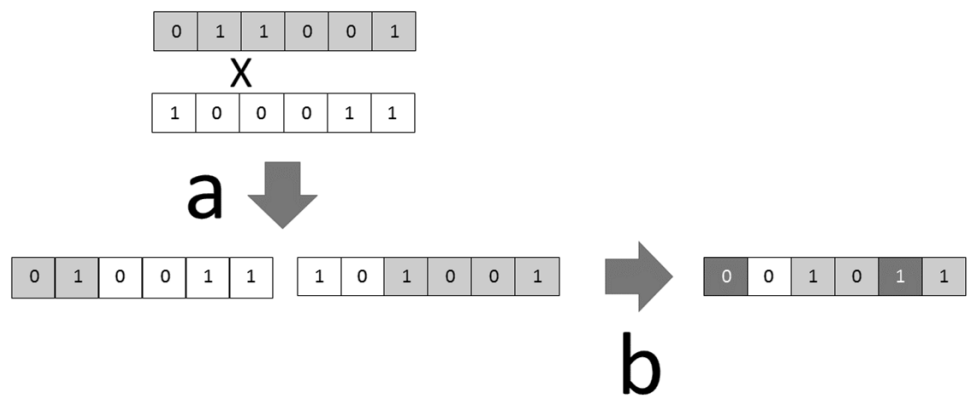

**Figure 2**. Genetic operators in chromosomes defined as a binary string. (a) One-point recombination. Once two parental chromosomes have been chosen, a point where the genes intersect, for example at the second position, is randomly selected. The chromosome segments are then exchanged, resulting in two recombinant chromosomes, i.e. the offspring. Note how the offspring chromosomes share the information of the parental chromosomes. (b) Mutation. In one of the child chromosomes, specifically the right one, we choose two positions at random, the first and the fifth in the example, inverting the value of the binary number.

Since the year 2000, several evolutionary strategies have been proposed to solve the Schrödinger equation. For example, [13] was able to solve the Schrodinger equation by means of genetic programming techniques. Applying this technique and if we know the power energy $V(x)$, it is possible to find a function and $E$-value that fit $V(x)$. Genetic programming is a particular case of genetic algorithm with the peculiarity of representing the functions or mathematical expressions as a tree. In particular, [14] were able to obtain an approximate solution of the Schrödinger equation by applying an approach known as grammatical evolution, thus a technique related to genetic programming whose output is a computer program. In 1988 Koza [15] introduces genetic programming and the application of this technique in optimization and search problems. Whether with a genetic algorithm or with genetic programming, the general protocol and the goal are always the same: to search for the optimal solution. In the particular case of a one-dimensional and time-independent system, based on the Hamiltonian represented in (4), the goal is to find the wave function that minimizes the energy.

Based on GAs, [16, 17] proposed a general procedure to solve the Schrödinger equation directly, evaluating the candidate solutions according to the energy. In a similar work [18] introduces a procedure that also solves the Schrödinger equation relying on a microgenetic algorithm. The novelty is the method for evaluating the solutions, which is referred as random point evaluation method. One of [18] authors, solves the Schrödinger equation [19] using both a microgenetic algorithm to obtain the parameters of a perceptron-like artificial neural network with which to represent the wave function. Recently there has been intense research on the use of deep learning methods in quantum computing, applying [20] artificial neural networks to obtain almost exact solutions of the Schrödinger equation.

*2.1. Solving the Schrödinger equation in elementary quantum systems*

In (2) the Hamiltonian operator represents the sum of the energies present in the system and therefore the physical environment where a particle exists, being $T(x)$ the kinetic energy and $V(x)$ the potential energy. Therefore, if we know the Hamiltonian of a system we can study the time evolution of its quantum states. In the present paper, we assume the simplest case, i.e. we study the Schrödinger equation in (i) a one-dimensional system, and (ii) the wave function only depends on the position occupied by a particle and not on time. Henceforth, when referring to the genetic algorithm, we will write the wave function $|\psi\rangle$ as $\psi(x)$. Considering the above, we can write (2):

$$(T(x) + V(x))\psi(x) = E\psi(x) \quad (5)$$

However, in the Schrödinger equation the usual way of writing the Hamiltonian of the system is as follows:

$$\left(-\frac{\hbar^2}{2\mu}\frac{d^2}{dx^2} + V(x)\right)\psi(x) = E\psi(x) \quad (6)$$

In (6), we have replaced the kinetic energy term $T(x)$ by $-\frac{\hbar^2}{2\mu}\frac{d^2}{dx^2}$ where $\hbar$ represents Planck's constant and $\mu$ the mass. In order to simplify and make the expression (6) more manageable we will assume that in the kinetic energy terms $\hbar$ and $\mu$ are equal to one, simplifying the kinetic energy term to $-\frac{1}{2}\frac{d^2}{dx^2}$ as shown below:

$$\left(-\frac{1}{2}\frac{d^2}{dx^2} + V(x)\right)\psi(x) = E\psi(x) \quad (7)$$

In this paper, we will illustrate how with a simple genetic algorithm it is possible to obtain an approximate solution of the Schrödinger equation. The chosen cases are simple but not trivial, and as we noted above the studied cases are one-dimensional with the wave function independent of time *t*. Therefore, we will write the Schrödinger equation of the systems studied as follows:

$$-\frac{1}{2}\frac{d^2}{dx^2}\psi(x) + V(x)\psi(x) - E\psi(x) = 0 \quad (8)$$

In the following, we will solve equation (6) for the case where the particle is confined into the region $x \in [a,b]$. If we use the second order derivative formula then we will discretize the Schrödinger equation. Therefore, applying the numerical grid method to the kinetic energy term of equation (8) we will obtain the following expression:

$$\frac{d^2}{dx^2}\psi(x) = \psi(x-1) + \psi(x+1) - 2\psi(x) \quad (9)$$

Hereafter we will continue the explanation of the procedure focusing the explanation on its implementation by the genetic algorithm, and therefore we will write (9) as shown below:

$$D_{i,j}^2 = \widehat{\psi}(x_{i,j-1}) + \widehat{\psi}(x_{i,j+1}) - 2\widehat{\psi}(x_{i,j}) \quad (10)$$

In (10) $D_{i,j}^2$ represents the term $\frac{d^2}{dx^2}\psi(x)$ and $\widehat{\psi}(x_{i,j})$ the value of the wave function at position *j* of chromosome *i*. Thus, the chromosome is a vector featuring a candidate wave function solution. For instance, a "chromosome 1" is a vector which elements $\widehat{\psi}(x_{1,1}), \widehat{\psi}(x_{1,2}), \ldots, \widehat{\psi}(x_{1,j})$ are the values of the wave function for positions 1, 2... *j* on the *x*-axis of the coordinate system where we will plot the obtained wave function. Note how for implementation purposes (8) can be written in either of the two ways shown below:

$$-\frac{1}{2}D_{i,j}^2 + V(x_j)\widehat{\psi}(x_{i,j}) - E\widehat{\psi}(x_{i,j}) = 0, \quad \frac{1}{2}D_{i,j}^2 + \left(E - V(x_j)\right)\widehat{\psi}(x_{i,j}) = 0 \quad (11)$$

Likewise, the expression to the left of (11) can be expressed in formal language as follows:

$$(\widehat{H} - E)\psi(x) \quad (12)$$

According to [18] the fitness of the chromosome *i* or solution $\widehat{\psi}(x_{i,1}), \widehat{\psi}(x_{i,2}), \ldots, \widehat{\psi}(x_{i,j})$, is given by the function $F_i$:

$$F_i = e^{-Z_i} \quad (13)$$

In the fitness function (13), $Z(i)$ is equal to:

$$Z_i = \frac{\sum |(\widehat{H} - E)\psi(x)|^2}{\sum |\psi(x)|^2} \quad (14)$$

calculating the $Z_i$ term in the genetic algorithm with the following expression:

$$Z(i) = \frac{\sum_j^l D_{i,j}^2}{\sum_{a+1}^{b-1} \widehat{\psi}(x_{i,j})^2} \quad (15)$$

Note how in term (15) the numerator is calculated with the following formula:

$$D_{i,j} = \frac{D_{i,j}^2}{2} + \left(E - V(x_j)\right)\widehat{\psi}(x_{i,j}) \quad (16)$$

Thus, for calculation purposes (16) can also be written:

$$D_{i,j} = \frac{\widehat{\psi}(x_{i,j-1}) + \widehat{\psi}(x_{i,j+1}) - 2\widehat{\psi}(x_{i,j})}{2} + \left(E - V(x_j)\right)\widehat{\psi}(x_{i,j}) \quad (17)$$

Once the $Z(i)$ value has been calculated with (15), in the genetic algorithm we will obtain the fitness value of chromosome $i$:

$$F(i) = e^{-Z(i)} \quad (18)$$

In expression (15) $l$ is the *x*-axis length and $a$, $b$ the real lower and upper bounds respectively. Note that the optimization problem solved by the genetic algorithm is the minimization of the fitness function $F$ satisfying the Schrödinger equation.

2.1.1. Genetic algorithm implementation

First, we define the positions on the *x*-axis, i.e. $X_{i1}$, $X_{i2}$... $X_{ij}$. The GA involves the following steps (Figure 1):

Step 0 - Initialization of the population. Using a random number generator with continuous uniform distribution in the interval [a, b] we generate a population of size $N$ ($i=1$ ... $N$) whose individuals are the chromosomes of length $j$, the genes being the values of the wave function $\widehat{\psi}(x_{i,j})$ for each position $j$ on the *x*-axis.

Step 1 - Evaluation of the chromosomes. Using the fitness function (18) we obtain $F(i)$, i.e. the goodness or merit of each chromosome $i$.

Step 2 - Selection. The chromosomes that will pass to the next generation are selected according to their fitness value by applying the method of the wheel parents selection [10, 12].

Step 3 - Crossover. Once the next generation is obtained, the recombination between parents is conducted applying a crossover genetic operator. For this purpose, two parental individuals are chosen by applying the tournament method. That is, the first parental chromosome is selected between two chromosomes chosen at random from the population, selecting the one with the highest fitness. The second parental chromosome is chosen in a similar way. Applying the one-point crossover operator [12], two offspring chromosomes are obtained replacing the new chromosomes to the parental chromosomes (Figure 2). A value of $p_r$ recombination rate is set at the beginning of the simulation experiment.

Step 4 - Mutation. Given a certain population of size $N$, the value of the population mutation rate $p_m$ is set at the beginning of the simulation experiment. If a chromosome $i$ undergoes mutation then the probability that a site $j$ of chromosome $i$ will experience a random change in the value of the wave function $\hat{\psi}(x_{i,j})$ is given according to the value of the parameter or rate $p_s$ specified in advance (Figure 2).

Step 5 - Steps 1 to 4 are repeated over and over again until the convergence condition is satisfied, either that the value of the fitness $F(i)$ is greater than or equal to a certain threshold value $\theta$ or a maximum value of the number of generations is reached.

2.1.2. Simulation experiments

In this work, we show how to solve the Schrödinger equation by applying a simple GA. To this end, we apply the procedure described above (Section 2.1.1) in three classical quantum mechanical systems: particle in a 1-dimensional box, the quantum harmonic oscillator and a simple model of the hydrogen atom.

A particle in a one-dimensional box is a quantum mechanics model describing the horizontal motion of a particle confined within a region, a well of infinite depth, from which the particle cannot escape. Given an energy value $E$, the squared value of the wave function, i.e. $|\psi(x)|^2$, gives the probability of locating the particle at a certain position inside the box. Since the potential energy $V(x)$ is zero inside the box with dimension $L$ ($V(x)=0$, $0<x<L$) and $V(x)$ is infinite at the walls of the box ($V(x)=\infty$, x<0 or x>L) then the Schrödinger equation (7) will be denoted by the following expression:

$$-\frac{1}{2}\frac{d^2}{dx^2}\psi(x) = E\,\psi(x) \quad (19)$$

The approximate solution of equation (19), i.e. the wave function, was obtained by setting the following values for the parameters of the GA: population size $N=44$, recombination rate $p_r=0.65$ and mutation rate $p_m=0.2$ ($p_s=0.1$). The termination criteria was a maximum number of 3200 generations or a fitness value equal or greater than 0.87. For the quantum system the potential energy value was $V=0$ and $E=0.02$.

The second quantum system simulated is the quantum harmonic oscillator which allows the study of quantum oscillatory systems, e.g. the vibration of molecules. If the system to be studied is the vibration of a diatomic molecule then the classical idea of a constant spring between the two atoms is used. In such case the potential energy is $V(x) = \frac{1}{2}x^2$. Using this potential energy value in the Schrödinger equation, we obtain the equation for the harmonic oscillator:

$$\left(-\frac{1}{2}\frac{d^2}{dx^2} + \frac{1}{2}x^2\right)\psi(x) = E\,\psi(x) \quad (20)$$

Similar to the previous experiment the approximated solution of equation (20) was obtained with the GA by choosing the same parameters and termination criteria. Regarding the quantum system, the value of the energy $E$ was set to 0.5.

Finally, we will illustrate how it is possible with a simple GA to obtain the approximate solution of the Schrödinger equation for an elementary model of the hydrogen atom [21, 22]. The hydrogen atom consists of two particles, a proton located at the mass center of the system and an electron at a distance $r$ from the central proton. The Schrödinger equation of an electron around a proton can be separated into two functions: the so-called spherical harmonic function describing the position of the electron around the proton, and the radial function representing the distance of the electron from the proton. The radial equation of the hydrogen atom is given by the following expression (21):

$$\left(-\frac{\hbar^2}{2\mu r}\frac{d^2}{dr^2}r + \frac{\hbar^2}{2\mu}\frac{l(l+1)}{r^2} - \frac{e^2}{4\pi\varepsilon_0}\frac{1}{r}\right)\psi = E\,\psi \quad (21)$$

where the potential energy includes a term $\frac{e^2}{4\pi\varepsilon_0}\frac{1}{r}$ which corresponds to the Coulomb potential energy, i.e. the attraction between the electron and proton. Equation (21) is simplified as in the previous cases [23], obtaining a Schrödinger equation for the hydrogen atom that does not include physical constants, as shown below:

$$\left(-\frac{1}{2}\frac{d^2}{dx^2} + \frac{l(l+1)}{x^2} - \frac{2}{x}\right)\psi(x) = E\,\psi(x) \quad (22)$$

Note how in this case the potential energy term is $V(x) = \frac{l(l+1)}{x^2} - \frac{2}{x}$, where $l$ is the quantum number describing the shape of the region occupied by the electron. Unlike the previous cases, the wave function obtained by solving the Schrödinger equation corresponds to the atomic orbital. The solution of the equation of this elementary atom, i.e. the atomic orbital, was obtained with the GA choosing the same values of the parameters of the previous simulation experiments. In the present experiment, the quantum number $l$ was 1 and the energy value $E$ was equal to -0.5.

Regarding the implementation method, we will use the expression (17) in the three quantum systems, changing the value of the potential energy $V(x)$ in each experiment according to the simulated quantum system. In other words, we obtained the numerical solution of the Schrödinger equation by applying to the expression (8) the procedure described in this Section with the help of a GA. The approximate wave function was compared with the expected or theoretical wave function, which was obtained by applying the procedure described in [24].

*2.2. Quantum neuron*

During the last years, one of the most interesting issues of study in quantum computing has been to formulate a quantum version of an artificial neural network [25, 26]. However, while the McCulloch-Pitts quantum definition of a neuron has been a successful goal, it has not been easy to find a quantum-learning algorithm. The purpose of this kind of algorithm is to modify the values of the weights associated with the neural connections, and thus simulate learning, which will allow the neural network to recognize a given pattern [12]. Figure 3 shows the quantum neuron model, representing the quantum version of the classical McCulloch-Pitts model [27].

Let there be a quantum neuron with N inputs $|x_1\rangle, |x_2\rangle, ..., |x_N\rangle$ in which the input $|x_i\rangle$ is a qubit:

$$|0\rangle = \begin{pmatrix} 1 \\ 0 \end{pmatrix}, \quad |1\rangle = \begin{pmatrix} 0 \\ 1 \end{pmatrix} \quad (23)$$

Therefore, $|\psi\rangle$ is expressed as:

$$|\psi\rangle = \alpha|0\rangle + \beta|1\rangle = \begin{pmatrix} \alpha \\ \beta \end{pmatrix} \quad (24)$$

From a biological point of view, note that an input signal $|1\rangle$ could be interpreted as an excitatory signal, while a signal $|0\rangle$ would be an inhibitory input signal. Once a neuron receives *i* input signals, these are processed yielding to an output $|y\rangle$, which is given by the following expression:

$$|y\rangle = \widehat{F} \sum_{i=1}^{N} \widehat{w_i} |x_i\rangle \quad (25)$$

where $\widehat{F}$, $\widehat{w_i}$ and $|x_i\rangle$ are the activation function, the connection weights and the input signals respectively. Therefore, in the model under study the sum $\widehat{w_1}.|x_1\rangle + \widehat{w_2}.|x_2\rangle$ (referred to as net value) will be the argument of the activation function $\widehat{F}$, obtaining with this function the output value $|y\rangle$ of the neuron.

Although we will adopt for practical purposes the definition (25), it is also interesting to consider other ways of defining at a formal level a quantum model of an artificial neural network. In this alternative definition, the quantum neuron model is based on the known fact that given an initial state $|\psi(0)\rangle$ or input signal, the solution of the Schrödinger equation at time *t* is given by the linear mapping:

$$|\psi(t)\rangle = e^{-i\widehat{H}t} |\psi(0)\rangle \quad (26)$$

Despite its limitations, the quantum version of the artificial neuron based on the classical McCulloch-Pitts neuron model is the most feasible for simulation experiments. However, other approaches are possible. For example, [28] define the evolution of a quantum neural network as the Hamiltonian:

$$\widehat{H} = \sum_j \widehat{w_j} \Xi_j \quad (27)$$

representing $\Xi_j$ any quantum circuit or unitary quantum gate. In such a case we obtain:

$$|\psi(t)\rangle = e^{-i\sum_j w_j \Xi_j \delta t}|\psi(0)\rangle = \prod_j e^{-iw_j \Xi_j \delta t}|\psi(0)\rangle \quad (28)$$

In other words, for formal purposes a quantum artificial neural network is according to [28]:

$$|\psi(t)\rangle = e^{-iw_1 \Xi_1 \delta t} \cdot e^{-iw_2 \Xi_2 \delta t} \cdot \ldots \cdot e^{-iw_k \Xi_k \delta t}|\psi(0)\rangle \quad (29)$$

In the present experiment, we will adopt (26) definition and we will assume that the connection weights and the activation function are 2x2 matrices. In the McCulloch-Pitts model, the activation function is the Heaviside step function. Thus, once the inputs have been processed if the summation exceeds a certain threshold then the neuron fires (activation state): the output is 1, otherwise the output is 0 (the neuron is at resting state).

In our study, adopting as 'toy model' the quantum neuron introduced by [29] we will conduct the following simulation experiment. According to this model we will assume an elementary quantum neuron with only two inputs $|x_1\rangle, |x_2\rangle$ and the following activation function:

$$\hat{F} = \frac{1}{\sqrt{2}}\begin{pmatrix} 0 & 1 \\ 1 & -1 \end{pmatrix}\begin{pmatrix} \text{sgn}(.) & 0 \\ 0 & \text{sgn}(.) \end{pmatrix} \quad (30)$$

where sgn(.) is the sign function:

$$\text{sgn}(x) = \begin{cases} 1 & \text{for } x > 0 \\ -1 & \text{for } x < 0 \end{cases} \quad (31)$$

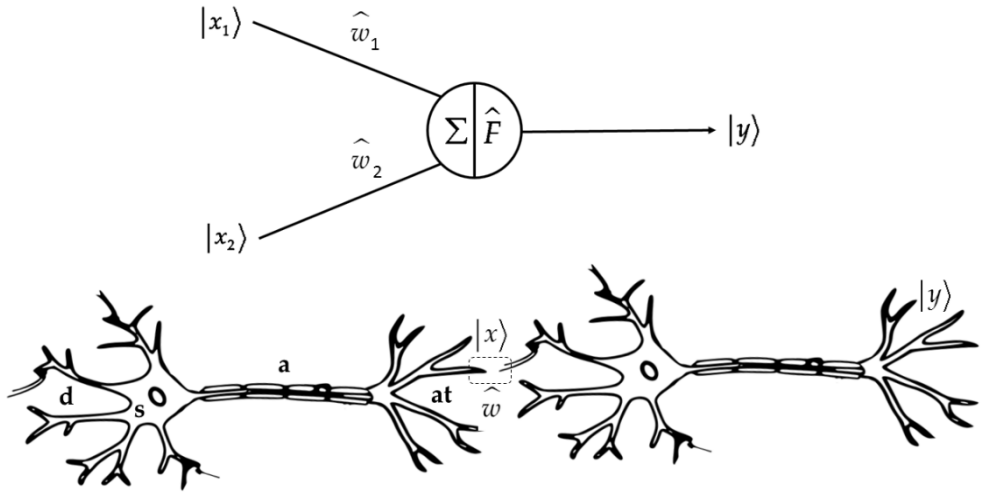

**Figure 3**. Quantum neuron. In the top section of the figure, is depicted the quantum version of the McCulloch-Pitts neuron. At the bottom of the figure, the neuron on the left displays its biological counterpart showing the dendrites (d), axon (a) and axonal terminations (at). The left neuron emits an input signal $|x\rangle$ to the right neuron that is modulated by the weight $\hat{w}$ associated with the synaptic connection between neurons (little box). Once the signal is processed by the right neuron, this last neuron emits an output or response $|y\rangle$. Learning involves a change in the value of the weights, modifying the response of the neuron on the right (for more details see text).

Likewise, in the model of quantum neuron [29] the weights $\widehat{w_1}$ and $\widehat{w_2}$ of the connections are given by the Hadamard gate:

$$H = \frac{1}{\sqrt{2}}\begin{pmatrix} 1 & 1 \\ 1 & -1 \end{pmatrix} \quad (32)$$

Consequently, the described model is the quantum version of a perceptron neural network implementing the truth table of the Boolean XOR operator (Table 1).

In quantum computing the quantum version of the perceptron rule is given by the following expression, which is similar to its non-quantum version:

$$\widehat{w}_i(t+1) = \widehat{w}_i(t) + \eta\left(|d\rangle - |y(t)\rangle\right)\langle x_i| \quad (33)$$

Note how in expression (33) a neural network 'learns', i.e. the weights are modified, according to whether the output $|y(t)\rangle$ of a neuron is the desired $|d\rangle$ or not, where $\eta$ is the learning rate. The problem arises from non-unitary outputs, i.e. gates output [30], during the training step of the neural network. This fact is a consequence of the operations performed during the adjustment of the connection weights.

**Table 1**. Quantum XOR operator

| $|x_1\rangle$ | $|x_2\rangle$ | $|y\rangle$ |
|---|---|---|
| $\begin{pmatrix} 1 \\ 0 \end{pmatrix}$ | $\begin{pmatrix} 1 \\ 0 \end{pmatrix}$ | $\begin{pmatrix} 1 \\ 0 \end{pmatrix}$ |
| $\begin{pmatrix} 0 \\ 1 \end{pmatrix}$ | $\begin{pmatrix} 1 \\ 0 \end{pmatrix}$ | $\begin{pmatrix} 0 \\ 1 \end{pmatrix}$ |
| $\begin{pmatrix} 1 \\ 0 \end{pmatrix}$ | $\begin{pmatrix} 0 \\ 1 \end{pmatrix}$ | $\begin{pmatrix} 0 \\ 1 \end{pmatrix}$ |
| $\begin{pmatrix} 0 \\ 1 \end{pmatrix}$ | $\begin{pmatrix} 0 \\ 1 \end{pmatrix}$ | $\begin{pmatrix} 1 \\ 0 \end{pmatrix}$ |

One of the applications of GAs has been the adjustment of the weights in a perceptron neural network. Inspired by this idea, we propose the following simulation experiment. We assume that learning in a quantum neuron, i.e. the adjustment of the weights $\widehat{w_1}$ and $\widehat{w_2}$, can be carried out evolutionarily with a GA. Adopting the quantum neuron model [29] described above, the goal of the present experiment was to find the appropriate weights for the neuron to implement the XOR operator. Therefore, a GA will search over several generations for the suitable values of weights $\widehat{w_1}, \widehat{w_2}$. In the simulation experiment, the possible values of the weights were as shown below:

$$w_0 = \frac{1}{\sqrt{2}}\begin{pmatrix} 1 & 1 \\ 1 & -1 \end{pmatrix}, \quad w_1 = \frac{1}{\sqrt{2}}\begin{pmatrix} 0 & 1 \\ 1 & 0 \end{pmatrix}, \quad w_2 = \frac{1}{\sqrt{2}}\begin{pmatrix} 1 & 0 \\ 0 & -1 \end{pmatrix}, \quad w_3 = \frac{1}{\sqrt{2}}\begin{pmatrix} 1 & 0 \\ 0 & 1 \end{pmatrix} \quad (34)$$

Note how the possible values $w_0, w_1, w_2$ and $w_3$ of the weights associated with the neuron connections are the Hadamard gate, the Pauli gates X and Z as well as the gate identity I, the latter multiplied by $1/\sqrt{2}$.

The fitness of the chromosomes was calculated as follows. Given a certain chromosome, the output of the neuron, i.e. the $|y(t)\rangle$ value, was obtained for each one of the possible input signal pairs. Figures 4-5 show the fitness of a chromosome for all possible values of the two weights $\widehat{w_1}, \widehat{w_2}$ as a function of the inputs $|x_1\rangle, |x_2\rangle$. At the implementation level, the output $|y(t)\rangle$ of the neuron results from the product of the activation function $\hat{F}$ by the net value:

$$|y(t)\rangle = \hat{F}.net \quad (35)$$

Afterwards, and once the output vector $|y(t)\rangle$ is already known, we obtain its transpose $|y(t)^T\rangle$. Next, we calculate the trace of the product between the output vector and its transpose. Finally, the fitness value of a neuron j is the sum of the traces corresponding to the four possible pairs of signals, i.e. $|0\rangle$ and $|0\rangle$, $|1\rangle$ and $|0\rangle$, $|0\rangle$ and $|1\rangle$, $|1\rangle$ and $|1\rangle$:

$$F(j) = \sum_4 \left| Tr\left(|y(t)\rangle |y(t)^T\rangle\right) \right| \quad (36)$$

| Inputs I1 I2 | W0 W0 | W1 W1 | W2 W2 | W3 W3 | W0 W1 | W0 W2 | W0 W3 | W1 W0 | W2 W0 | W3 W0 |
|---|---|---|---|---|---|---|---|---|---|---|
| $\binom{1}{0}\binom{1}{0}$ | $\binom{1}{0}$ | $\binom{1}{-1}$ | $\binom{0}{1}$ | $\binom{0}{1}$ | $\binom{1}{-½}$ | $\binom{½}{½}$ | $\binom{½}{½}$ | $\binom{1}{-½}$ | $\binom{½}{½}$ | $\binom{½}{½}$ |
| $\binom{0}{1}\binom{1}{0}$ | $\binom{0}{1}$ | $\binom{½}{0}$ | $\binom{½}{0}$ | $\binom{½}{0}$ | $\binom{0}{½}$ | $\binom{½}{½}$ | $\binom{½}{½}$ | $\binom{½}{½}$ | $\binom{0}{½}$ | $\binom{½}{-½}$ |
| $\binom{1}{0}\binom{0}{1}$ | $\binom{0}{1}$ | $\binom{½}{0}$ | $\binom{½}{0}$ | $\binom{½}{0}$ | $\binom{½}{½}$ | $\binom{0}{½}$ | $\binom{1}{-½}$ | $\binom{0}{½}$ | $\binom{½}{½}$ | $\binom{½}{½}$ |
| $\binom{0}{1}\binom{0}{1}$ | $\binom{1}{0}$ | $\binom{0}{1}$ | $\binom{1}{-1}$ | $\binom{1}{-1}$ | $\binom{½}{½}$ | $\binom{1}{-½}$ | $\binom{0}{½}$ | $\binom{½}{½}$ | $\binom{1}{-½}$ | $\binom{0}{½}$ |
| Fitness | 4 | 1 | 1 | 1 | 1 | 1 | 1 | 1 | 1 | 1 |

**Figure 4**. Chromosome fitness according to the input and the value of the connection weights.

| Inputs I1 I2 | W1 W2 | W1 W3 | W2 W3 | W2 W1 | W3 W1 | W3 W2 |
|---|---|---|---|---|---|---|
| $\begin{pmatrix}1\\0\end{pmatrix}\begin{pmatrix}1\\0\end{pmatrix}$ | $\begin{pmatrix}½\\0\end{pmatrix}$ | $\begin{pmatrix}½\\0\end{pmatrix}$ | $\begin{pmatrix}0\\1\end{pmatrix}$ | $\begin{pmatrix}½\\0\end{pmatrix}$ | $\begin{pmatrix}½\\0\end{pmatrix}$ | $\begin{pmatrix}0\\1\end{pmatrix}$ |
| $\begin{pmatrix}0\\1\end{pmatrix}\begin{pmatrix}1\\0\end{pmatrix}$ | $\begin{pmatrix}0\\1\end{pmatrix}$ | $\begin{pmatrix}0\\1\end{pmatrix}$ | $\begin{pmatrix}½\\0\end{pmatrix}$ | $\begin{pmatrix}0\\0\end{pmatrix}$ | $\begin{pmatrix}1\\-1\end{pmatrix}$ | $\begin{pmatrix}½\\0\end{pmatrix}$ |
| $\begin{pmatrix}1\\0\end{pmatrix}\begin{pmatrix}0\\1\end{pmatrix}$ | $\begin{pmatrix}0\\0\end{pmatrix}$ | $\begin{pmatrix}1\\-1\end{pmatrix}$ | $\begin{pmatrix}½\\0\end{pmatrix}$ | $\begin{pmatrix}0\\1\end{pmatrix}$ | $\begin{pmatrix}0\\1\end{pmatrix}$ | $\begin{pmatrix}½\\0\end{pmatrix}$ |
| $\begin{pmatrix}0\\1\end{pmatrix}\begin{pmatrix}0\\1\end{pmatrix}$ | $\begin{pmatrix}½\\0\end{pmatrix}$ | $\begin{pmatrix}½\\0\end{pmatrix}$ | $\begin{pmatrix}0\\0\end{pmatrix}$ | $\begin{pmatrix}½\\0\end{pmatrix}$ | $\begin{pmatrix}½\\0\end{pmatrix}$ | $\begin{pmatrix}0\\0\end{pmatrix}$ |
| Fitness | 1 | 1 | 1 | 1 | 1 | 1 |

**Figure 5**. Chromosome fitness according to the input and the value of the connection weights (continuation).

2.2.1. Genetic algorithm implementation

The simulation experiments were performed with $N=12$ chromosomes with two genes which values are the weights $\widehat{w_1}$ and $\widehat{w_2}$ of the connections. In the present experiment, because of the short length of the chromosomes, we do not apply the crossover operator in the GA. Therefore, the GA uses only the mutation operator as a source of variability, with $p_m=0.6$ and $p_s=0.1$. The GA included the following steps:

Step 0 - Initialization of the population. Using a random number generator we generate a population of size $N$ ($j=1 \ldots N$) whose individuals or chromosomes comprise the values of the weights $\widehat{w_1}$ and $\widehat{w_2}$, i.e. the $\frac{1}{\sqrt{2}} U_{2x2}$ matrices with {-1, 0, 1} elements. In the present experiment, the allowed values of the weights are restricted to $w_0$, $w_1$, $w_2$ or $w_3$ (34).

Step 1 - Evaluation of the chromosomes. Using the fitness function (36) we obtain with $F(j)$ the goodness or merit of each chromosome.

Step 2 - Selection. The chromosomes that will pass to the next generation are selected according to their fitness value by applying the wheel parents selection method [10, 12].

Step 3 - Mutation. Given a certain population of size $N$, the values of the parameters $p_m$ and $p_s$ rate are set at the beginning of the simulation experiment.

2.2.2. Simulation experiments

In contrast to the other experiments described in this paper, in the present experiment we limit ourselves to search with the GA for the suitable weights. The search ends when the output of the quantum neuron and the two input signals matches the XOR operator. Obviously and according to the problem space (Figures 4-5) the GA goes through the evolutionary surface in search of the optimal solution, which corresponds to the values $w_0$ and $w_0$ of the two connections present in the neuron.

*2.3. Evolving quantum circuits for robots*

The 1990s of the previous century were the golden age of research on autonomous agents [31] as part of the Artificial Life [32] projects that were then being developed. Eventually some of the Artificial Life models were rescued and implemented in their quantum version. For example, [33] designed quantum Braitenberg vehicles, i.e. Lego robots with "quantum brains" controlled by multiple-valued quantum circuits simulated via software. A Braitenberg vehicle [34] is an autonomous agent that exhibits reactive behaviors, e.g. movement, in response to input signals. In the experiments, the most frequent signal was the light detected by its sensors. (Figure 6). The senso-motor architecture [12] of these quantum robots is hybrid, thus the control device is a quantum circuit whereas its sensors and effectors are classical devices. Therefore, according to [33] the behavior exhibited by the robots is based on the principles of quantum mechanics, i.e. entanglement, superposition and collapse of the wave function (measurement).

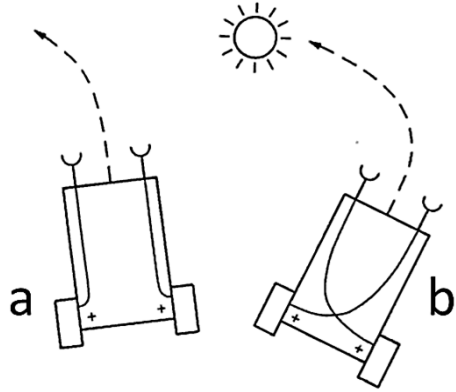

**Figure 6**. Braitenberg vehicles with different wiring (a, b) between the light input signal and the output (wheel motors).

In this paper, based on [33] model we show how it is possible to synthesize with an elementary GA the quantum circuit or "brain" that will control the behavior of a quantum Braitenberg robot. For some initial state $|\psi(0)\rangle$, i.e. the signal detected by a sensor, the Hamiltonian operator represents the sequence of operations performed in a quantum system, i.e. the quantum circuit implementing the control device. In consequence, given an initial state $|\psi(0)\rangle$ or *input* signal, the solution of the Schrödinger equation at time *t* is given by the following linear mapping in the state space:

$$|\psi(t)\rangle = U(t)|\psi(0)\rangle \quad (37)$$

It is important to note that at theoretical level $U(t)$ represents the solution of the Schrödinger equation. Once the quantum circuit or "brain" processes the *input*, the *output* is the behavior exhibited by the robot via its effectors. Consequently, the output results from measuring or observing the qubit returned by the quantum circuit and therefore it is due to the collapse of the wave function $|\psi\rangle$.

At the practical level, the operator *U* can be decomposed into *q* elementary quantum gates, designing the quantum circuitry that controls the quantum Braitenberg robot [33] with elementary quantum gates (Figure 7). Note how in this experiment, the solution of the Schrödinger equation is implemented in the quantum circuit responsible for the behavior exhibited by the robot in response to a given stimulus (Figure 6).

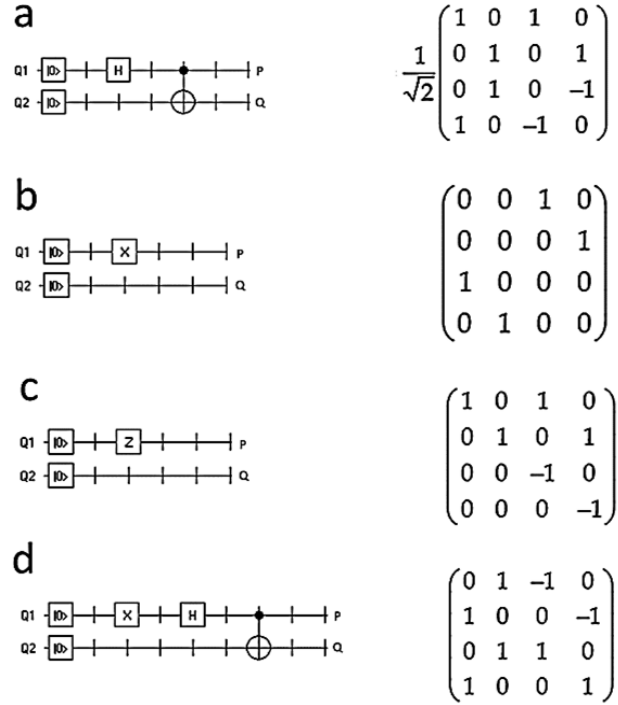

**Figure 7**. Quantum circuits and their matrix representation. A circuit plays the role of an 'artificial brain' able to control the behavior of a Braitenberg robot depending on the light signal received or not by the sensors Q1 and Q2 (for explanation see text).

The goal of the present experiment was to synthesize with a GA different control circuits in order to program different behaviors in the Braitenberg robots. In general, the GA is similar to the previous one with the exception of the details related with chromosome definition, fitness function, etc. According to Figure 7, four simulation experiments were performed synthesizing in each case with the GA the appropriate control quantum circuit. The circuit $i$ to be synthesized is referred as matrix $U_i$ whereas the target circuit is represented by a matrix $U_t$, with the latter matrix accounting for each one the circuits shown in Figure 7.

Matrices $U_t$, i.e. the matrices representing the desired quantum circuits, were obtained as explained next. The elements of a quantum circuit, i.e. wires and quantum gates, are unitary operators represented by unitary matrices (Table 2). If two elements are in parallel, then the Kronecker product is applied between the circuit matrices. However, if two elements are connected in series, then the circuit matrices are multiplied in the standard way (inner product). For instance, in the circuit represented in Figure 7a, first we solve $H$ and $I$:

$$H \otimes I = \frac{1}{\sqrt{2}} \begin{pmatrix} 1 & 1 \\ -1 & 1 \end{pmatrix} \otimes \begin{pmatrix} 1 & 0 \\ 0 & 1 \end{pmatrix} = \frac{1}{\sqrt{2}} \begin{pmatrix} 1 & 0 & 1 & 0 \\ 0 & 1 & 0 & 1 \\ 1 & 0 & -1 & 0 \\ 0 & 1 & 0 & -1 \end{pmatrix} \quad (38)$$

**Table 2.** Quantum gates used in the circuits of the Braitenberg robot experiments, their symbol and their unitary matrix.

| Quantum gate | Symbol | Unitary matrix |
|---|---|---|
| Wire | ———— | $I = \begin{pmatrix} 1 & 0 \\ 0 & 1 \end{pmatrix}$ |
| H (Hadamard) | —[H]— | $H = \dfrac{1}{\sqrt{2}} \begin{pmatrix} 1 & 1 \\ 1 & -1 \end{pmatrix}$ |
| Z (Pauli-Z) | —[Z]— | $Z = \begin{pmatrix} 0 & 1 \\ 0 & -1 \end{pmatrix}$ |
| X (Pauli-X) | —[X]— | $X = \begin{pmatrix} 0 & 1 \\ 1 & 0 \end{pmatrix}$ |
| CNOT | ⊕ | $CNOT = \begin{pmatrix} 1 & 0 & 0 & 0 \\ 0 & 1 & 0 & 0 \\ 0 & 0 & 0 & 1 \\ 0 & 0 & 1 & 0 \end{pmatrix}$ |

Secondly, we multiply the result obtained previously with the CNOT gate:

$$(H \otimes I).CNOT = \frac{1}{\sqrt{2}} \begin{pmatrix} 1 & 0 & 1 & 0 \\ 0 & 1 & 0 & 1 \\ 1 & 0 & -1 & 0 \\ 0 & 1 & 0 & -1 \end{pmatrix} \cdot \begin{pmatrix} 1 & 0 & 0 & 0 \\ 0 & 1 & 0 & 0 \\ 0 & 0 & 0 & 1 \\ 0 & 0 & 1 & 0 \end{pmatrix} = \frac{1}{\sqrt{2}} \begin{pmatrix} 1 & 0 & 1 & 0 \\ 0 & 1 & 0 & 1 \\ 0 & 1 & 0 & -1 \\ 1 & 0 & -1 & 0 \end{pmatrix} \quad (39)$$

Once we set the unitary matrix $U_t$ by the procedure described above, the GA will obtain the fitness value of chromosome or matrix $U_i$:

$$F(i) = 1 - \frac{1}{2^n} \left| Tr\,(U_i.U_t^{-1}) \right| \quad (40)$$

where $n$ is the number of qubits. Note that the maximum fitness is obtained once the objective function or fitness is minimized. When the GA find the right circuit, the product then $U_i.U_t^{-1}$ is the identity matrix $I$. Therefore with $n=2$ and the trace of the matrix $I$ equal to 4, $F(i)$ will be equal to one in the experiments conducted with the GA. However, the procedure described and used in [35] is somewhat cumbersome, proposing in this paper an alternative procedure in which the fitness value $F(i)$ is calculated more quickly. Our procedure obtains the fitness value by calculating the Hamming distance between the $U_i$ and $U_t$ matrices by comparing the rows of both matrices:

$$F(i) = U_i \oplus U_t \quad (41)$$

2.3.1. Genetic algorithm implementation

The simulation experiments were performed with $N=200$ being the termination criterion a maximum number of generations equal to 1000. Note how in this experiment the candidate chromosomes or solutions were not vectors but matrices. The recombination rate was $p_r =0.6$, whereas with respect to the mutation rate $p_m$ different values (0.1, 0.2, 0.3 and 0.4) were studied setting the value $p_s =0.1$. Therefore, and in accordance with the above parameters, the GA comprises the following steps:

Step 0 - Initialization of the population. Using a random number generator we generate a population of size $N$ ($i=1 ... N$) whose individuals or chromosomes are $U_{4x4}$ matrices with {-1, 0, 1} elements.

Step 1 - Evaluation of the chromosomes. Using the fitness function (41) we obtain with $F(i)$ the goodness or merit of each matrix $U_i$.

Step 2 - Selection. The chromosomes, i.e. matrices $U$, which will pass to the next generation, are selected according to their fitness value by applying the method of wheel parents selection [10, 12].

Step 3 - Crossover. Two parental matrices are chosen by applying the tournament method described above (Section 2.1.1), obtaining the recombinant chromosomes by means of the one-point crossver operator (Figure 2). A value of $p_r$ recombination rate is set at the beginning of the simulation experiment.

Step 4 - Mutation. Given a certain population of size $N$, the value of the population mutation rate $p_m$ is set at the beginning of the simulation experiment. If a matrix $U_i$ undergoes mutation then the probability that an element of the matrix will experience a random change in its value, e.g. from -1 to 1, is decided according to the value of the $p_s$ rate previously set.

2.3.2. Simulation experiments

In this work, we perform four rounds of experiments, showing in each round how a simple GA can synthesize one of the four quantum circuits shown in Figure 7, thus the hardware of the "brain" governing the behavior of a Braitenberg vehicle. Once a quantum circuit is synthesized controlling the behavior of a robot, we assume the existence of an external light source that can be *on* or *off*. Light will be the input signal received by the robot through the appropriate sensor. In the experiments, we define the states of the input signal based on two qubits as follows: if input is $|00\rangle$ then the light source is off, whereas if input is $|11\rangle$ then the light source is on. The response or output of the robot, i.e. its behavior, will be *stop*, *left turn*, *right turn*, and *forward motion* when the output is $|00\rangle$, $|01\rangle$, $|10\rangle$ and $|11\rangle$ respectively.

**3. Results**

The results obtained and presented in this section show how a GA is a useful heuristic procedure in the search for an approximate solution of the Schrödinger equation. In an elementary quantum system (Section 2.1.2) and using expression (17) within a GA, we conclude that GA usefulness is presumably similar to traditional numerical methods. However, in complex quantum systems or in quantum artificial intelligence experiments whose solution involves the treatment of the Schrödinger equation, such is the case of the 'toy models' studied in this work (Sections 2.2-2.3), GAs represent a very useful tool. In the latter case, we illustrate with two simulation experiments how a GA is able to find the proper solutions. For instance, in a first simulation experiment, with the help of a GA we found the appropriate weights in an elementary quantum neural circuit (Section 2.2). In consequence, given a certain initial state or *input* $|\psi(0)\rangle$ and based on expression (26), we obtain the appropriate or expected *output*. Thus, the gate corresponding to an XOR operator. Likewise, in a second simulation experiment and according to the expression (37) it is shown how a GA is a suitable procedure to synthesize the quantum circuit $U(t)$ or 'brain' that will drive the behavior of a Braitenberg vehicle.

*3.1. Approximate solution of the Schrodinger equation in three elementary quantum systems*

The results obtained show how a standard GA is able to provide an approximate solution of the Schrödinger equation, at least in the three elementary quantum systems that we have considered in this work. Figures 8-10a display the performance graph of the GA, showing how the average fitness increases over the generations, and therefore the convergence of the population of chromosomes to the optimal solution. Note that the population is the set of the *i* chromosomes or solutions $\hat{\psi}(x_{i,1}), \hat{\psi}(x_{i,2}), \ldots, \hat{\psi}(x_{i,j})$, each chromosome being a vector with the values obtained from the wave function with respect to *x*-axis. The plot of the approximate solution of the Schrödinger equation, i.e. the wave function obtained with the GA, is also displayed in Figures 8-10b. The results for each elementary quantum system were as follows.

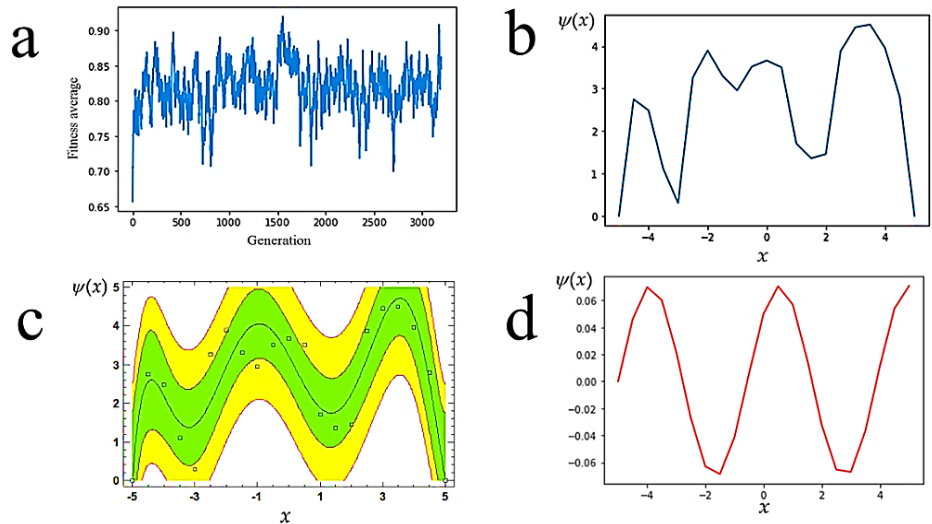

**Figure 8**. Experiment of the particle confined in a one-dimensional box. (a) Performance graph of the GA. (b) Approximate wave function obtained with the GA. (c) Polynomial function fitted to the approximate wave function. (d) Theoretical wave function.

In the experiment of the particle confined in a one-dimensional box as well as in the other two experiments, we fit the wave function points obtained with the GA to a polynomial function (Figures 8-10c). Polynomial regression was performed in all the cases studied with the statistical package STATGRAPHICS Centurion 18, version 18.1.12. In the experiment of the particle in a one-dimensional box with a *p*-value equal to 0.0012 the relation between $\psi(x)$ and $x$ (Figure 8c) was given by the function:

$$\psi(x) = 3.2221 - 1.5008x - 0.2908x^2 + 0.4434x^3 + 0.0387x^4 - 0.0306x^5 - 0.0012x^6 + 0.0006x^7 \quad (42)$$

According to the $R^2$ value, the fitted model would explain the 79.5099% of the variability of the wave function $\psi(x)$.

Figure 9 shows in the model of the harmonic oscillator the performance graph of the GA as well as the approximate wave function, its polynomial regression fitting and finally the expected or theoretical wave function. In the harmonic oscillator the wave function values obtained with the GA were fitted with a *p*-value equal to 0.0000 to a fourth order polynomial model (Figure 9c), the fitted model equation being:

$$\psi(x) = 4.1085 - 0.0147x - 0.5521x^2 + 0.0010x^3 + 0.0164x^4 \quad (43)$$

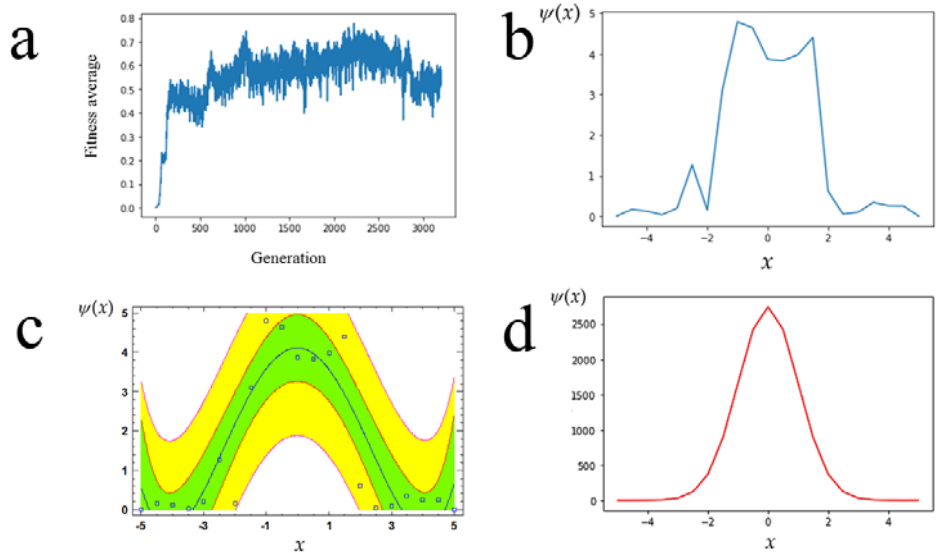

**Figure 9**. Experiment of the harmonic oscillator. (a) Performance graph of the GA. (b) Approximate wave function obtained with the GA. (c) Polynomial function fitted to the approximate wave function. (d) Theoretical wave function.

The $R^2$ statistic indicates that the fitted model explains a 79.0245% of the $\psi(x)$ variability.

Finally, the radial Schrödinger equation of the hydrogen atom was also solved obtaining the approximate wave function with the GA (Figure 10), fitting (*p*-value=0.0221) the obtained values of the wave function to a fourth order polynomial model (Figure 10c):

$$\psi(x) = 0.0103 + 0.0642x - 0.0138x^2 + 0.0010x^3 - 0.000027x^4 \quad (44)$$

In the elementary atom model, the $R^2$ statistic stated that the fitted model explains the 49.1172% of the $\psi(x)$ variability.

In each of the elementary quantum systems that we have studied, the theoretical or expected wave function was also obtained (Figures 8-10d). The theoretical wave function was obtained by solving with a program in Python language the normalize wave function:

$$\psi(x) = \frac{\psi(x)}{\sqrt{\int_{-x\max}^{x\max} \psi(x)^2 dx}} \quad (45)$$

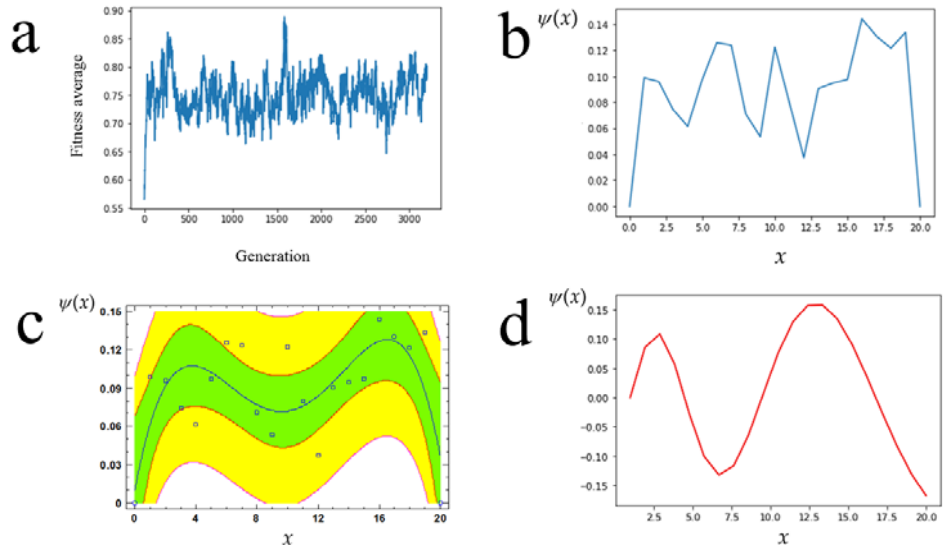

**Figure 10**. Experiment of the hydrogen atom. (a) Performance graph of the GA. (b) Approximate wave function obtained with the GA. (c) Polynomial function fitted to the approximate wave function. (d) Theoretical wave function.

*3.2. Evolutionary adjustment of the connection weights of a quantum neuron*

The conducted experiments show how a GA is able to fit the weights associated with the connections by finding in very few generations (Figure 11) the optimal values of $w_1$ and $w_2$. Figure 11 shows the execution graph of the GA plotting the average fitness value per generation. Thus, we found with a GA a quantum neuron with the ability to process the input signals as an XOR operator. In consequence, an evolutionary algorithm, i.e. a procedure that solves an optimization problem by means of an approach inspired by the Darwinian notion of natural selection [12], can be successfully applied to a quantum neural network. Using a GA, we conducted the learning step of the neural network. One of the advantages of applying this kind of heuristic techniques is that we avoid getting non-unitary outputs during network training. The reason is because the fitness function would penalize non-unitary solutions and these flawed 'species' would disappear along the evolutionary process.

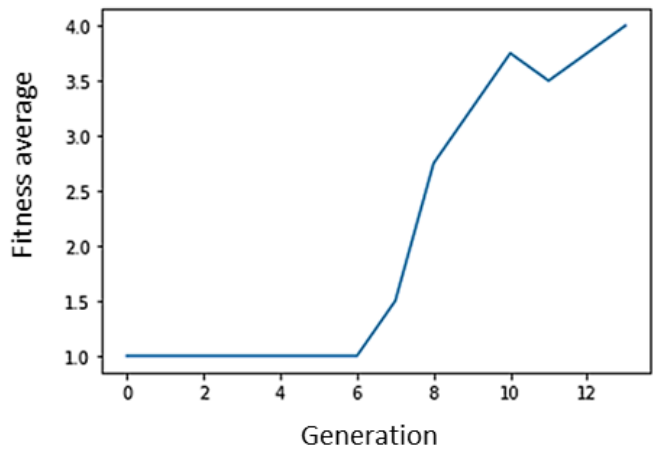

**Figure 11**. Performance graph of the GA used in the learning experiment (adjustment of weights) of a quantum neuron.

*3.3. Synthesis of a quantum circuit controlling the behavior of a Braitenberg robot.*

The simulation results show the feasibility of synthesizing a quantum circuit by means of a GA. In the present experiments the chromosomes are the solutions $U_i$ of the matrix $U_t$ standing for a desired quantum circuit. Figure 12 shows the execution graph of the GA plotting the maximum fitness value, i.e. the chromosome with the best solution, in each generation.

Table 3 shows the generation number in which the execution of the GA ends because the GA found the optimal solution, i.e. the desired quantum circuit that will control the behavior of the Braitenberg vehicle. Statistical analysis of Table 3 with a Chi-square test of independence resulted in a *p*-value below 2.2e-16. Therefore, we can conclude that the generation number at which the GA finds the optimal quantum circuit depends on both the circuit type (Figure 7) and the mutation rate $p_m$.

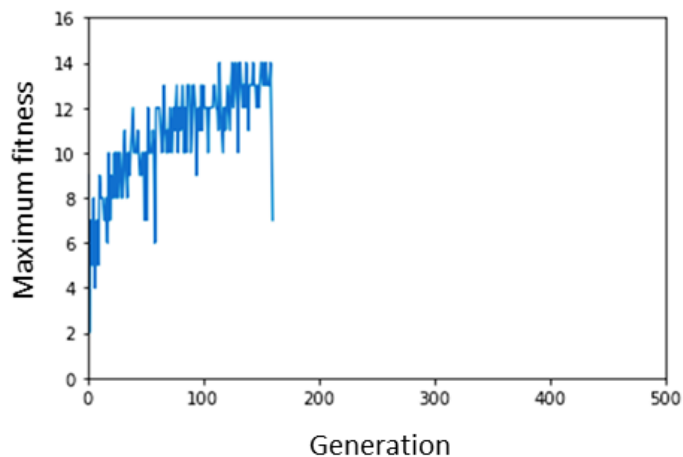

**Figure 12**. Performance graph of the GA which is used to design the quantum circuit controlling the behavior of a Braitenberg robot.

**Table 3.** Contingency table for the number of generations needed in a GA to find a valid solution as a function of circuit type and mutation rate.

| Quantum circuit | $p_m$=0.1 | $p_m$=0.2 | $p_m$=0.3 | $p_m$=0.4 |
|---|---|---|---|---|
| Figure 7a | 161 | 171 | 268 | 334 |
| Figure 7b | 213 | 121 | 101 | 141 |
| Figure 7c | 133 | 47 | 121 | 105 |
| Figure 7d | 96 | 183 | 58 | 137 |

**Table 4.** Braitenberg behavior.

| Quantum circuit | Light off $\|00\rangle$ | Light on $\|11\rangle$ |
|---|---|---|
| Figure 7a | *stop* $\|00\rangle$, *forward motion* $\|11\rangle$ | *turn left* $\|01\rangle$, *turn right* $\|10\rangle$ |
| Figure 7b | *turn right* $\|10\rangle$ | *turn left* $\|01\rangle$ |
| Figure 7c | *stop* $\|00\rangle$ | *turn left* $\|01\rangle$, *forward motion* $\|11\rangle$ |
| Figure 7d | *turn left* $\|01\rangle$, *forward motion* $\|11\rangle$ | *turn left* $\|01\rangle$, *forward motion* $\|11\rangle$ |

Once a quantum circuit is obtained, we proceeded to test the behavior of a Braitenberg robot in the presence $|11\rangle$ and absence $|00\rangle$ of a light signal. Table 4 shows the predictable behaviors of the robot as a function of the quantum circuit responsible for processing the light signal. Next, we show the general procedure to find out the behavior of the robot in each of the different simulated scenarios of Table 4.

Suppose the Einstein-Podolsky-Rosen (EPR) circuit shown in Figure 7a controls the behavior of a Braitenberg vehicle. What happens in the absence of light? That is, if the light signal is off then both inputs Q1 and Q2 of the circuit will be equal to $|0\rangle$, therefore:

$$|00\rangle = |0\rangle \otimes |0\rangle = \begin{pmatrix} 1 \\ 0 \end{pmatrix} \otimes \begin{pmatrix} 1 \\ 0 \end{pmatrix} = \begin{pmatrix} 1 \\ 0 \\ 0 \\ 0 \end{pmatrix} \quad (46)$$

When the EPR circuit processes the input, the result will be the quantum state $|\psi_i\rangle$:

$$|\psi_i\rangle = \frac{1}{\sqrt{2}} \begin{pmatrix} 1 & 0 & 1 & 0 \\ 0 & 1 & 0 & 1 \\ 0 & 1 & 0 & -1 \\ 1 & 0 & -1 & 0 \end{pmatrix} \begin{pmatrix} 1 \\ 0 \\ 0 \\ 0 \end{pmatrix} = \frac{1}{\sqrt{2}} \begin{pmatrix} 1 \\ 0 \\ 0 \\ 1 \end{pmatrix} \quad (47)$$

and thus:

$$|\psi_i\rangle = \frac{1}{\sqrt{2}}|00\rangle + \frac{1}{\sqrt{2}}|11\rangle \quad (48)$$

Indeed, there will be a 50% probability that the robot is stopped $|00\rangle$ and another 50% that it is moving forward $|11\rangle$.

Now, we will consider another possible situation. In agreement with Table 4 if the robot is provided with the control circuit with the Pauli X gate (Figure 7b) then in a dark environment $|00\rangle$ the robot would be turning continuously to the right $|10\rangle$. However, when a light signal $|11\rangle$ is emitted then the robot reverses its motion by turning to the left $|01\rangle$. Thus, if a light signal is emitted, in the input Q1 and Q2 of the circuit we get:

$$|11\rangle = |1\rangle \otimes |1\rangle = \begin{pmatrix} 0 \\ 1 \end{pmatrix} \otimes \begin{pmatrix} 0 \\ 1 \end{pmatrix} = \begin{pmatrix} 0 \\ 0 \\ 0 \\ 1 \end{pmatrix} \quad (49)$$

Once the quantum circuit processes the input signal, we obtain the following quantum state:

$$|\psi_i\rangle = \begin{pmatrix} 0 & 0 & 1 & 0 \\ 0 & 0 & 0 & 1 \\ 1 & 0 & 0 & 0 \\ 0 & 1 & 0 & 0 \end{pmatrix} \begin{pmatrix} 0 \\ 0 \\ 0 \\ 1 \end{pmatrix} = \begin{pmatrix} 0 \\ 1 \\ 0 \\ 0 \end{pmatrix} \quad (50)$$

It is interesting to note how the quantum state at the microscopic level results in an 'observable behavior' at the macroscopic level. At a theoretical level, this fact occurs once an observer measures or observes the quantum state of the Braitenberg vehicle, collapsing the wave function. In the present scenario, if we calculate the resulting probability:

$$|\langle \psi_i | \psi_i \rangle|^2 = (0)(0) + (1)(1) + (0)(0) + (0)(0) = 1 \quad (51)$$

and normalize the state dividing by $\sqrt{1}$, then we obtain the same result. In other words, the result corresponds to the base vector $|01\rangle$ and the probability of observing this behavior is 100%.

Now assume that the robot was equipped with the circuit depicted in Figure 7c. In such a case and according to Table 4, in an environment with the light signal off the robot would be stopped. Once the light signal is switched on, signal reaches the Q1 and Q2 inputs of the quantum circuit. Afterwards the signal is processed by the quantum circuit, resulting the following output:

$$|\psi_i\rangle = \begin{pmatrix} 1 & 0 & 1 & 0 \\ 0 & 1 & 0 & 1 \\ 0 & 0 & -1 & 0 \\ 0 & 0 & 0 & -1 \end{pmatrix} \begin{pmatrix} 0 \\ 0 \\ 0 \\ 1 \end{pmatrix} = \begin{pmatrix} 0 \\ 1 \\ 0 \\ -1 \end{pmatrix} \quad (52)$$

In this case, to measure the quantum state we calculate the corresponding probability:

$$|\langle \psi_i | \psi_i \rangle|^2 = (0)(0) + (1)(1) + (0)(0) + (-1)(-1) = 2 \quad (53)$$

Next, if we normalize the quantum state dividing by $\sqrt{2}$, we obtain:

$$|\psi_i\rangle = \frac{1}{\sqrt{2}}\begin{pmatrix} 0 \\ 1 \\ 0 \\ -1 \end{pmatrix} \quad (54)$$

Finally, we will calculate the amplitude probability of the obtained quantum state, collapsing into one of the four basis states $|00\rangle, |01\rangle, |10\rangle$ or $|11\rangle$:

$$\langle 00|\psi\rangle = \frac{1}{\sqrt{2}}(1\ 0\ 0\ 0)\begin{pmatrix} 0 \\ 1 \\ 0 \\ -1 \end{pmatrix} = 0 \ , \ \langle 01|\psi\rangle = \frac{1}{\sqrt{2}}(0\ 1\ 0\ 0)\begin{pmatrix} 0 \\ 1 \\ 0 \\ -1 \end{pmatrix} = \frac{1}{\sqrt{2}}$$

$$\langle 10|\psi\rangle = \frac{1}{\sqrt{2}}(0\ 0\ 1\ 0)\begin{pmatrix} 0 \\ 1 \\ 0 \\ -1 \end{pmatrix} = 0 \ , \ \langle 11|\psi\rangle = \frac{1}{\sqrt{2}}(0\ 0\ 0\ 1)\begin{pmatrix} 0 \\ 1 \\ 0 \\ -1 \end{pmatrix} = -\frac{1}{\sqrt{2}} \quad (55)$$

Consequently, we can conclude that once the Braitenberg vehicle detects the light signal, the probability of the robot turning left $|01\rangle$ or moving forward $|11\rangle$ is 50% for each one of these behaviors.

**4. Discussion**

In this paper, we show how GAs are very useful heuristic methods in those practical situations where the approximate solution of the Schrödinger equation is implicitly required. In general, when the stationary solution of an elementary but non-trivial quantum system is needed within the experimental protocol then the application of a standard GA can successfully address this step. In this case, all that will be required is to define a population of chromosomes represented by strings of amplitude probabilities where each amplitude plays the role of a gene. For example, [16] using classical genetic operators to perform the selection, crossover and mutation shows how it is possible to obtain approximate solutions of the Schrödinger equation in three elementary quantum systems. These quantum systems were the symmetric double well problem, the H atom model and the two-dimensional coupled harmonic oscillator problem. Therefore, in this work we corroborate how a simple GA is sufficient to obtain an approximate solution of the Schrödinger equation, including for its usefulness the Python language code of the GA described in Section 2.1.1, customized for the experiments described in Section 2.1.2.

In the field of the electronic structure of atoms [36] explore different strategies of GAs with which to solve approximately the Schrödinger equation from the Hamiltonian of a system composed of "N electrons and M nuclei". Thus, in different classes of molecules. In this review they describe three methods, which they call variational-recipe, direct solution and density-based approach, whose application they illustrate with different ions. However, the direct solution method with a standard GA [37] applied to solve the Schrödinger equation is the standard protocol of choice for researchers. Using this protocol, there are studies whose authors propose versions of GAs with some methodological features. Thus, given the GA nature the possibility of parallelizing a GA is very useful in those cases in which it is necessary to compute the energy of the wave function in each one of the generations. For instance, in [17] their authors illustrate an example in which using a cluster under the Linux operating system, they solve the Schrödinger equation for a 2D harmonic oscillator interacting via Hooke law forces. In this experiment, they calculate the kinetic energy with the Fourier transform and the potential energy by numerical integration. Other protocols involve modifications of a standard GA in order to avoid premature convergence and local trapping of the population in a local optimum; such is the case of the microgenetic algorithm described in [18]. Moreover, other sophisticated approaches [19] obtain approximate solutions of the Schrödinger equation by combining the use of a GA with a perceptron-type artificial neural network [12]. This kind of proposals has led to the possibility of solving the Schrödinger equation by replacing the GA by an artificial neural network.

Recently [38] applied a deep neural network denoted PauliNet in quantum chemistry. The network was able to learn the complex patterns of the electrons located around the nuclei of a molecule. The success in the application of artificial neural networks in solving the Schrödinger equation has led to the study of the usefulness of other machine learning methods in quantum chemistry [39].

However, the application of GAs in quantum mechanics and quantum computing is not a novel methodology. According to [40] one of the most important limitations of quantum computation is the loss of fidelity of quantum operations. In [40], their authors propose the use of a GA to suppress errors in quantum simulation experiments conducted in digital environments. In addition, GAs have been successfully used for the synthesis of quantum circuits [41, 42]. In the present work, this possibility has also been demonstrated in the experiments we have carried out with quantum robots, i.e. with Braitenberg vehicles. In this case, the objective or fitness function is based on the comparison of the candidate matrix representing the quantum circuit evolved with the GA, with the expected matrix or target circuit. An interesting feature of [41] is the use of GAs including classical mutation operators as well as other less common genetic operators. For instance, insertion, deletion operators inspired by genetic mechanisms present in organisms.

In [42] it is shown how it is possible synthesizing quantum circuits equivalent to Boolean circuits, including the simulation of the evolution of an oracle. The synthesis of oracles with GAs is an extremely useful task in quantum computing. That is, given as input $n$ qubits, a circuit conducting a unitary transformation $U$ of the qubits leading to certain output can be obtained evolutionarily with a GA.

It is obvious that the synthesis of quantum circuits via GAs is an optimization problem, just because of the possibility of decomposing a unitary matrix into a sequence of quantum gates. Once again, and in the field of quantum circuit design, the problem of the premature convergence of the GA is a frequent problem. In order to solve this particular disadvantage, [43] proposes the use of a GA based on the Island Model, a solution reminiscent of the one adopted by [17].

In the field of nanoelectronics, GAs and other evolutionary algorithms have been applied to the synthesis of circuits built with quantum-dot cellular automata (QCA). In [44] we performed the implementation of the XOR gate through the evolutionary synthesis of a QCA circuit. This experiment would be bearing some resemblance to the one performed in the present work with a quantum neuron. It is interesting to note that making use of a simulator, in our case QCADesigner [45], the Schrödinger equation was used by the simulator engine to obtain the state of a QCA automaton with respect to those of its neighborhood in the circuit. Indeed, this is the case because a QCA automaton is a system with two states, being its Hamiltonian is given by:

$$H_i = \sum_j \begin{pmatrix} -\frac{1}{2} P_j E_{i,j}^k & -\gamma_i \\ -\gamma_i & \frac{1}{2} P_j E_{i,j}^k \end{pmatrix} \quad (56)$$

where $E_{i,j}^k$ is the kink energy between cells *i* and *j*, $P_j$ is the cell polarization and $\gamma$ is the tunneling energy.

The Python source code of the experiments performed with the three elementary quantum systems (Section 2.1), the quantum neuron circuit (Section 2.2) and the Braitenberg vehicles (Section 2.3) is hosted at [46].

In this paper, we have made a review of a set of different scenarios in which the solution of the Schrödinger equation is involved or required, showing how GAs are heuristic methods that can satisfactorily substitute the use of numerical methods.

**Acknowledgments:** I am very grateful to Professor Manuel Alfonseca, at the Computer Science School of the Universidad Autónoma de Madrid, for his invaluable collaboration during the interpretation and implementation of the genetic algorithm, and for the time he took to review some of the details of this article.

## References

1. Chow, P.C. Computer solutions to the Schrödinger equation. *American Journal of Physics* **1972**, *40*, 730-734.

2. Senese, F. *Symbolic Mathematics for Chemists*, 1st ed.; Wiley, 2019; pp. 323–324.

3. Binesh, A.; Mowlavi, A.A.; Arabshahi, H. Application of Runge-Kutta numerical methods to solve the Schrödinger equation for hydrogen and positronium atoms. *IJRRAS* **2010**, *5*(*3*), 289-293.

4. Jorgensen, L.; Lopes Cardozo, D.; Thibierge, E. Numercial resolution of the Schrödinger equation. École Normale Supérieure de Lyon, Master Sciences de la Matière 2011, Numerical Analysis Project. Available online:

5. Matthew, N.S.; Upadhyay, S.; Madura, J.D. A python program for solving Schrödinger equation in undergraduate physical chemisitry. Journal of Chemical Education **2017**, *94*, 813-815.

6. Obi Tayo, B. Finite difference solution of the Schrödinger equation. Modern Physics. 2019. Available online: https://medium.com/modern-physics/finite-difference-solution-of-the-schrodinger-equation-c49039d161a8

7. Figueiras, E.; Olivieri, D.; Paredes, A.; Michinel, H. An open source virtual laboratory for the Schrödinger equation. *European Journal of Physics* **2018**, *39* 055802. https://doi.org/10.1088/1361-6404/aac999

8. Duran-Avendaño, O.; Ramirez-Martín, C. Solución de la ecuación de Schrödinger mediante LabVIEW. *Av. cien. ing.* **2012**, *3*(*4*), 177-184.

9. Ulmann, B. Solving the Schrödinger equation. *Analog Computer Applications* **2019**, Issue 21, 2*1*-MAY. https://analogparadigm.com/documentation.html

10. Goldberg, D.E. *Genetic Algorithms in Search, Optimization, and Machine Learning*; Addison-Wesley: Reading, Massachusetts, 1989.

11. Davis, L (Ed.). *Handbook of Genetic Algorithms*; Van Nostrand Reynhold: New York, USA, 1991.

12. Lahoz-Beltra, R. *Bioinformática: Simulación, Vida Artificial e Inteligencia Artificial*. Transl.: Spanish. Ediciones Díaz de Santos: Madrid, Spain, 2004; pp. 237–323.

13. Makarov, D.E.; Metiu, H. Using genetic programming to solve the Schrödinger equation. *J. Phys. Chem. A* **2000**, *104,* 8540-8545.

14. Jebari, K.; Madiafi, M.; Elmoujahid, A. An evolutionary approach for solving Schrödinger equation. *arXiv:1402.5428v1* **2014**. https://doi.org/10.48550/arXiv.1402.5428

15. Koza, J.R. *Genetic Programming: On the Programming of Computers by Means of Natural Selection (Complex Adaptive Systems)*; A Bradford Book, The MIT Press, 1992.

16. Saha, R.; Chaudhury, P.; Bhattacharyya, S.P. Direct solution of Schrödinger equation by genetic algorithm: tests cases. *Physics Letters A* **2001**, *291*, 397-406.

17. Saha, R.; Bhattacharyya, S.P.; Taylor, C.D.; Zhao, Y.; Cundari, T.R. Direct solution of the Schrödinger equation by a parallel genetic algorithm: Cases of an exactly solvable 2-D interacting oscillator and the hydrogen atom. *International Journal of Quantum Chemistry* **2003**, *94*, 243-250.

18. Nakanishi, H.; Sugawara, M. Numerical solution of the Schrödinger equation by a microgenetic algorithm. *Chemical Physics Letters* **2000**, *327*, 429-438.

19. Sugawara, M. Numerical solution of the Schrödinger equation by neural network and genetic algorithm. *Computer Physics Communications* **2001**, *140*, 366-380.

20. Hermann, J.; Schätzle, Z.; Noé, F. Deep neural network solution of the Schrödinger equation. *Nat Chem.* **2020**, *12*(*10*), 891-897.

21. The Schrödinger Wave Equation for the Hydrogen Atom. Available online: 4.10.1 https://chem.libretexts.org/@go/page/20298 (accessed on 7-7-2022).

22. Ertl, M.C. Solving the stationary one dimensional Schrödinger equation with the shooting method. Bachelor Thesis, Faculty of Electrical Engineering and Information Technology Institute for Microelectronics, TU Wien, September 2016. Available online: https://www.iue.tuwien.ac.at/uploads/tx_sbdownloader/Bachelor-Arbeit_Marie_ERTL_09-2016.pdf

23. Finite Difference Method for Solving Schrödinger Equation. Bucknell University. Available online: http://linux.bucknell.edu/~mligare/python_projects/quantum/finiteDifference_hydrogen.html (accessed on 7-7-2022).

24. Rioux, F. Introduction to Numerical Solutions of Schödinger's Equation. Available online: https://chem.libretexts.org/@go/page/135863 (accessed on 7-7-2022).

25. Zhang, Y.; Ni, Q. Design of quantum neuron model for quantum neural networks. *Quantum Engineering* **2021**, 3:e75. https://doi.org/10.1002/que2.75

26. Tacchino, F.; Macchiavello, C.; Gerace, D.; Bajoni, D. An artificial neuron implemented on an actual quantum processor. *npj Quantum Inf* **2019**, *5(26)*. https://doi.org/10.1038

27. McCulloch, W.S.; Pitts, W. A logical calculus of the ideas immanent in nervous activity. *Bulletin of Mathematical Biophysics* **1943**, *5*, 115–133.

28. Dendukuri, A.; Keeling, B.; Fereidouni, A.; Burbridge, J.; Luu, K.; Churchill, H. Defining quantum neural networks via quantum time evolution. *arXiv:1905.10912v2* **2020**. https://doi.org/10.48550/arXiv.1905.10912

29. Nayak, S.; Nayak, S; Singh, J.P. Computational power of quantum artificial neural network. *International Journal of Computer Science and Technology* **2011**, *2*(*2*), 35-37.

30. De Paula Neto, F.M.; de Oliveira, W.R. Analysis of quantum neural models. In 11º Congresso Brasileiro de Inteligência Computacional - CBIC 2013, Porto de Galinhas, PE, September 2013, pp. 1-7.

31. Maes, P. Artificial life meets entertainment: lifelike autonomous agents. *Commun. ACM* **1995**, *38*(*11*), 108–114. https://doi.org/10.1145/219717.219808

32. Langton, C.G (Ed). *Artificial Life: An overview*, 5th ed.; A Bradford Book, The MIT Press, 2000.

33. Raghuvanshi, A.; Fan, Y.; Woyke, M.; Perkowski, M. Quantum Robots for Teenagers. In 37th International Symposium on Multiple-Valued Logic (ISMVL'07), 2007, pp. 18-18. doi: 10.1109/ISMVL.2007.46.

34. Braitenberg, V. *Vehicles: Experiments in synthetic psychology*; Cambridge, MA: MIT Press, 1984.

35. Miranda, F.T.; Paulo Balbi, P.; C.S. Costa, P. Synthesis of quantum circuits with an island genetic algorithm. *arXiv:2106.03115v1* **2021**. https://doi.org/10.48550/arXiv.2106.03115

36. Sharma, R.; Saha, R.; Nandy, S.; Prasad Bhattacharyya, S.; Chaudhury, P. Computation of molecular electronic structure by genetic algorithm. *Materials and Manufacturing Processes* **2009**, *24(2)*, 155-161.

37. Sarkar, K.; Prasad Bhattacharyya, S. *Soft-Computing in Physical and Chemical Sciences: A Shift in Computing Paradigm*. CRC Press: Boca Raton, FL, 2018; pp. 160-165.

38. Hermann, J.; Schätzle, Z.; Noé, F. Deep neural network solution of the electronic Schrödinger equation. *Nature chemistry* **2020**, *12*(*10*), 891–897.

39. Manzhos, S. Machine learning for the solution of the Schrödinger equation. *Mach. Learn.: Sci. Technol.* **2020**, *1*(*1*), 013002. https://doi.org/10.1088/2632-2153/ab7d30

40. Las Heras, U.; Alvarez-Rodriguez, U.; Solano, E.; Sanz, M. Genetic algorithms for digital quantum simulations. *Physical Review Letters* **2016**, 116, 230504.

41. Mukherjee, D.; Chakrabarti, A.; Bhattacherjee, D. Synthesis of quantum circuits using genetic algorithm. *International Journal of Recent Trends in Engineering* **2009**, *2*(*1*), 212-216.

42. Satsangi, S.; Gulati, A.; Kumar Kalra, P.; Patvardhan, C. Application of genetic algorithms for evolution of quantum equivalents of Boolean circuits. *International Journal of Nuclear and Quantum Engineering* **2012**, *6*(*3*), 275 - 279.

43. Miranda, F.T.; Baldi, P.P.; Costa, P.C.S. Synthesis of quantum circuits with an island genetic algorithm. *arXiv:2106.03115v1* **2021**. https://doi.org/10.48550/arXiv.2106.03115

44. Salas Machado, R.; Castellanos, J.; Lahoz-Beltra, R. Evolutionary synthesis of QCA circuits: A critique of evolutionary search methods based on the Hamming oracle. *International Journal "Information Technologies & Knowledge"* **2016**, *10*(*3*), 203-215.

45. Appendix A. Tutorial on QCADesigner 2.0.3. In: Sasamal, T.N.; Singh, A.K.; Mohan, A. *Quantum-Dot Cellular Automata Based Digital Logic Circuits: A Design Perspective*. Studies in Computational Intelligence 879, Springer Nature: 2020. Available online: https://link.springer.com/content/pdf/bbm:978-981-15-1823-2/1.pdf

46. Lahoz-Beltra, R. Solving the Schrödinger equation with genetic algorithms. Figshare Softw. 2022. https://doi.org/10.6084/m9.figshare.21210206.v1